\documentclass[%
 reprint,
superscriptaddress,
 amsmath,amssymb,
 aps,
pra,
]{revtex4-2}

\usepackage{graphicx}% Include figure files
\usepackage{dcolumn}% Align table columns on decimal point
\usepackage{bm}% bold math
\usepackage{hyperref}% add hypertext capabilities
\usepackage{subfigure}
\usepackage{physics}
\usepackage{algorithm}
\usepackage{algorithmic}
\usepackage{braket}
\usepackage{color,soul} % For highlighitng
\usepackage{nicefrac}
\begin{document}

\preprint{APS/123-QED}

\title{A saturation-absorption rubidium magnetometer with multilevel optical Bloch-equation modeling for intermediate-to-high fields}% Force line breaks with \\

\author{Mayand Dangi}
\email{mdangi2@wisc.edu}
\affiliation{Department of Electrical and Computer Engineering,
University of Wisconsin–Madison, Madison, Wisconsin 53706, USA}

\author{Prateek Rajan Gupta}
\affiliation{Department of Physics,
University of Wisconsin–Madison, Madison, Wisconsin 53706, USA}

\author{Walter Van Dyke}
\affiliation{Department of Nuclear Engineering and Engineering Physics, University of Wisconsin–Madison, Madison, Wisconsin 53706, USA}

\author{Joseph Kasti}
\affiliation{Department of Electrical and Computer Engineering,
University of Wisconsin–Madison, Madison, Wisconsin 53706, USA}

\author{Nivedan Vishwanath}
\affiliation{Department of Nuclear Engineering and Engineering Physics, University of Wisconsin–Madison, Madison, Wisconsin 53706, USA}

\author{Michael Zepp}
\affiliation{Department of Nuclear Engineering and Engineering Physics, University of Wisconsin–Madison, Madison, Wisconsin 53706, USA}

\author{David Smith}
\affiliation{Department of Nuclear Engineering and Engineering Physics, University of Wisconsin–Madison, Madison, Wisconsin 53706, USA}

\author{Benedikt Geiger}
\affiliation{Department of Nuclear Engineering and Engineering Physics, University of Wisconsin–Madison, Madison, Wisconsin 53706, USA}

\author{Jennifer T. Choy}
\email{jennifer.choy@wisc.edu}
\affiliation{Department of Electrical and Computer Engineering,
University of Wisconsin–Madison, Madison, Wisconsin 53706, USA}
\affiliation{Department of Physics,
University of Wisconsin–Madison, Madison, Wisconsin 53706, USA}

%  \email{Second.Author@institution.edu}
% \affiliation{%
%  Authors' institution and/or address\\
%  This line break forced with \textbackslash\textbackslash
% }%

% \collaboration{MUSO Collaboration}%\noaffiliation

%  \homepage{http://www.Second.institution.edu/~Charlie.Author}
% \affiliation{
%  Second institution and/or address\\
%  This line break forced% with \\
% }%
% \affiliation{
%  Third institution, the second for Charlie Author
% \affiliation{%
%  Authors' institution and/or address\\
%  This line break forced with \textbackslash\textbackslash
% }%

% \collaboration{CLEO Collaboration}%\noaffiliation

\date{\today}% It is always \today, today,

\begin{abstract}
We present SASHMAG (Saturated Absorption Spectroscopy High-field MAGnetometer), an atomic sensor based on Rubidium-87 ($^{87}\text{Rb}$) operating in the hyperfine Paschen-Back regime, demonstrating absolute magnetic field retrieval from $0.2\,\text{T}$ to $0.4\,\text{T}$ and a sensitivity of $139\,\mu\text{T}/\sqrt{\text{Hz}}$ at $\sim 0.4 \, \text{T}$, limited by laser scan rate and frequency calibration uncertainty. The all-optical sensor head, containing only a vapor cell and passive optics, is designed for deployment in environments with strong electromagnetic interference and ionizing radiation. To interpret the sub-Doppler spectra in this strongly field-dependent regime, we develop a comprehensive multilevel optical Bloch-equation model formulated explicitly in the uncoupled $\ket{m_I, m_J}$ basis, capturing state mixing and nonlinear saturation dynamics. This model reproduces measured spectra at sub-Doppler resolution and is independently validated against analytical predictions for power broadening and thermal Doppler scaling. Field estimation is performed by minimizing the residual between extracted line centers and transition frequencies calculated from the field-dependent Hamiltonian. The validated simulation establishes a foundation for synthetic-dataset generation toward autonomous magnetometry in applications ranging from MRI to fusion reactors.
% \begin{description}
% Secondary publications and information retrieval purposes.
% You may use the \texttt{description} environment to structure your abstract;
% use the optional argument of the \verb+\item+ command to give the category of each item. 
% \end{description}
\end{abstract}

\maketitle

%\tableofcontents

%%%%%%%%%%%%%%%%%%%%%%%%%%%%%%%%%%%%%%%%%%%%%%%%%%%%%%%%%%%%%%%%%%%%%%%%%%%%

\section{\label{sec:level1}Introduction}
Intermediate-to-high magnetic fields, spanning the sub-tesla regime to tens of tesla, are central to applications including ultra-high-field magnetic resonance imaging (MRI) \cite{kraff2015mri, trattnig2018key, okada2022neuroimaging}, condensed-matter and quantum materials studies \cite{zapf2014bose, ong_quantum_2021, lewin2023review, kartsovnik2004high}, and large-scale magnet systems for particle colliders \cite{bottura2022superconducting, shen2022design, muon2021magnetic} and nuclear fusion \cite{vayakis_development_2012, ma_design_2016, moreau_new_2018}. Nuclear fusion is among the most demanding: tokamak and stellarator superconducting magnets generate toroidal fields ranging from sub-tesla to several tesla \cite{vayakis_development_2012, moreau_new_2018}, and the shift from pulsed to steady-state operation places additional demands on magnetic diagnostics \cite{boivin2016diagnostics, ma_design_2016, bolshakova2012hall}. Precise knowledge of these fields is essential for plasma equilibrium reconstruction, magnetohydrodynamic (MHD) stability control, and machine protection \cite{vayakis_development_2012}, yet sensors in this environment must operate reliably under temperature variations, electromagnetic interference, neutron radiation, and strong fields \cite{quercia2022longterm, bolshakova2012hall}.

Existing magnetic sensors carry limitations in this space. Hall-effect probes offer wide dynamic range, compactness, and vector capability \cite{lakeshore_teslameter}, but their absolute accuracy is limited by thermal drift, susceptibility to electromagnetic interference, and radiation-induced changes in carrier concentration \cite{bolshakova2012hall, quercia2022longterm}. The ITER magnetic diagnostic design, for example, identifies the long-term stability of Hall sensors under cumulative neutron fluence as a leading open issue \cite{vayakis_development_2012, ma_design_2016}. 
Fluxmeter inductive coils are sensitive to changes in magnetic flux but require an external reference for absolute fields and suffer integration drift over long discharges \cite{buzio2011fabrication, moreau_new_2018}. Magneto-optical Faraday rotation magnetometers provide an alternative, but quantitative measurements depend on material-specific Verdet constants with strong wavelength and temperature dependence, requiring careful calibration and environmental control \cite{tsujilio2001fiberoptic}. Diamond nitrogen-vacancy (NV) magnetometry has recently been extended  $>1\,\text{T}$ as a complementary quantum platform with potential for radiation hardness \cite{graham2026high}, using microwave readout and per-sample calibration of the diamond's spin Hamiltonian.

Vapor-cell alkali magnetometry offers an attractive route to absolute field measurement because field-dependent line splitting is tied to atomic constants rather than phenomenological calibration factors, but existing approaches face tradeoffs between field range, spectral resolution, and sensor geometry that limit their utility in the intermediate-to-high field regime. Pulsed-field experiments have demonstrated alkali spectroscopy at fields up to $\sim60$ T using Doppler-limited absorption in microsized samples \cite{george2017pulsed, ciampini2017optical} and Zeeman-split emission lines on pulsed-power platforms \cite{garn1966technique, gomez2014magnetic, banasek2016measuring1, banasek2016measuring2}, establishing high-field atomic spectroscopy as a viable diagnostic but with resolution constrained by the pulsed magnetic environment. 
Nanocell-based work \cite{klinger2020proof, sargsyan2015study} has achieved sub-Doppler resolution at fields up to $\sim 0.2$ T, with extension above $0.5$ T projected, using sub-wavelength interaction lengths that yield weak single-pass absorption and require specialized cell fabrication. More recently, sideband-overlap Zeeman spectroscopy in a MEMS Cs vapor cell has demonstrated tesla-scale field metrology at a fixed operating point \cite{guo2026tesla}. Fixed-field Cs spectroscopy in MRI environments \cite{staerkind2023precision, staerkind2024high} achieves excellent sensitivity but is optimized for tracking a single transition at a known field strength.

In this work, we present SASHMAG (Saturated Absorption Spectroscopy High-field MAGnetometer), a Rubidium-87 ($^{87}$Rb) sensor that addresses these tradeoffs in the intermediate-to-high field regime. We demonstrate absolute field retrieval from 0.2 to 0.4 T with $\pm 561 \,\mu\text{T}$ precision and a sensitivity of $139\,\mu\text{T}/\sqrt{\text{Hz}}$ at $\sim 0.4 \, \text{T}$ with a projected sensitivity floor of $8\,\mu\text{T}/\sqrt{\text{Hz}}$ achievable through faster scanning and improved frequency calibration, using sub-Doppler spectroscopy on a bulk vapor cell with an all-optical sensor head. To interpret the resulting spectra quantitatively, we develop a comprehensive multilevel optical Bloch equation (OBE) model formulated explicitly in the uncoupled $|m_I, m_J\rangle$ basis appropriate for the hyperfine Paschen-Back regime.  Existing tools such as ElecSus accurately model Doppler-broadened linear absorption for absolute field extraction \cite{zentile2015elecsus, keaveney2018elecsus}, but do not capture the saturation dynamics or sub-Doppler features that emerge in high-field saturation absorption spectroscopy.
Our OBE framework is complementary to recent saturation-regime models of single-beam alkali absorption in the HPB regime \cite{haupl2025modelling}; we model the counter-propagating SAS lineshape specifically, reproducing measured spectra across the measurement range with independent validation against analytic predictions for power broadening and Doppler thermal scaling. By capturing saturation dynamics and state mixing in the regime where they matter, the model provides a foundation for generating synthetic training datasets for machine-learning-based field estimation in future autonomous sensing systems.

\section{\label{sec:level2}All-Optical Magnetometery Using Rubidium Vapor}
SASHMAG infers the magnetic field from the saturated absorption spectrum (SAS) of the $^{87}$Rb $\mathrm{D}_2$ line, which in the hyperfine Paschen-Back regime decomposes into a set of well-resolved Zeeman transitions whose frequencies depend monotonically on the applied field. This section describes the experimental setup, presents the measured high-field spectra, and demonstrates quantitative magnetic-field estimation with $\mu$T-level precision.

% The deployable sensor head will contain only passive optical components and an alkali vapor cell - no electronics, no metallic conductors - with all active components located remotely, giving the route for EMI-free magnetic sensing. 

\subsection{\label{ssec:expt_setup}Experimental Setup}

\begin{figure}[t]
\includegraphics[width=0.45\textwidth, trim={2.705cm 0 2.16cm 0 0}, clip]{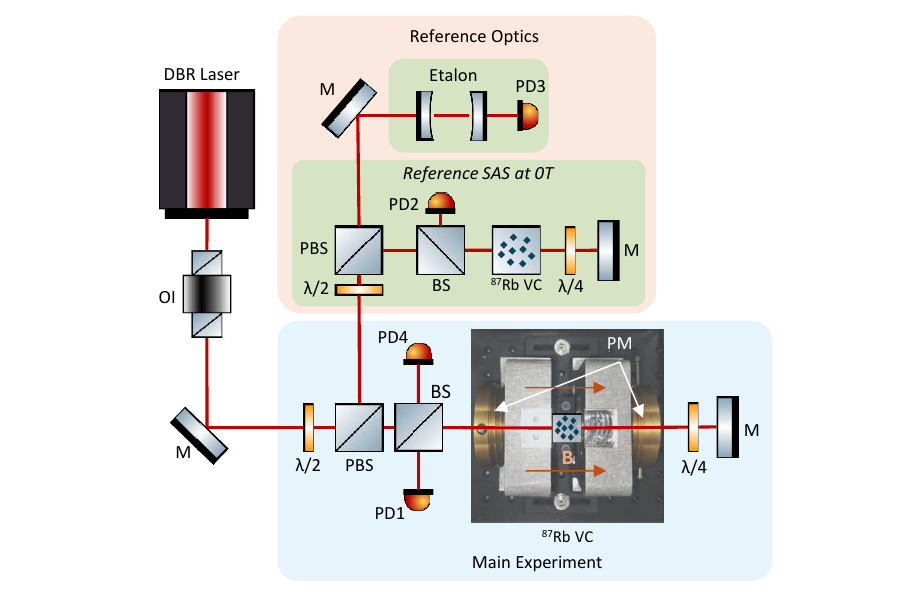}
\caption{\label{fig:optical_setup0} Optical schematic of the experimental setup for saturated absorption spectroscopy (SAS) of $^{87}\text{Rb}$. 
% The light from the DBR Laser is split into two arms. The reference optics arm (top, orange section) provides coarse frequency markers using an Etalon and a stable, zero-field ($0\text{ T}$) SAS lock signal using a separate ${}^{87}\text{Rb}$ cell and $\mathbf{PD2}$ for frequency positioning. The main experiment arm (bottom, blue section) performs the high magnetic field SAS measurement. This arm features the primary ${}^{87}\text{Rb}$ vapor cell ($\mathbf{VC}$) placed in a Faraday geometry between two permanent magnets ($\mathbf{PM}$) configured in a quasi-Helmholtz arrangement. The field-dependent spectrum is detected at $\mathbf{PD1}$. 
Key components include: \textbf{VC} (Vapor Cell), \textbf{PM} (Permanent Magnets), \textbf{OI} (Optical Isolator), \textbf{M} (Mirror), $\lambda/2$ (Half Waveplate) and $\lambda/4$ (Quarter Waveplate), \textbf{PBS} (Polarizing Beam Splitter), \textbf{BS} (Beam Splitter), and \textbf{PD} (Photodiode).}
\end{figure}

The experimental setup is illustrated in Fig. \ref{fig:optical_setup0}, utilizing two counter-propagating pump and probe beams. The optical setup comprises two distinct sections: Reference optics for frequency calibration and a main experiment arm for high magnetic field measurements.

\begin{figure*}[t]
\centering
% {left bottom right top}
\includegraphics[width=\textwidth, trim={0.2cm 0.2cm 2.6cm 0cm}, clip]{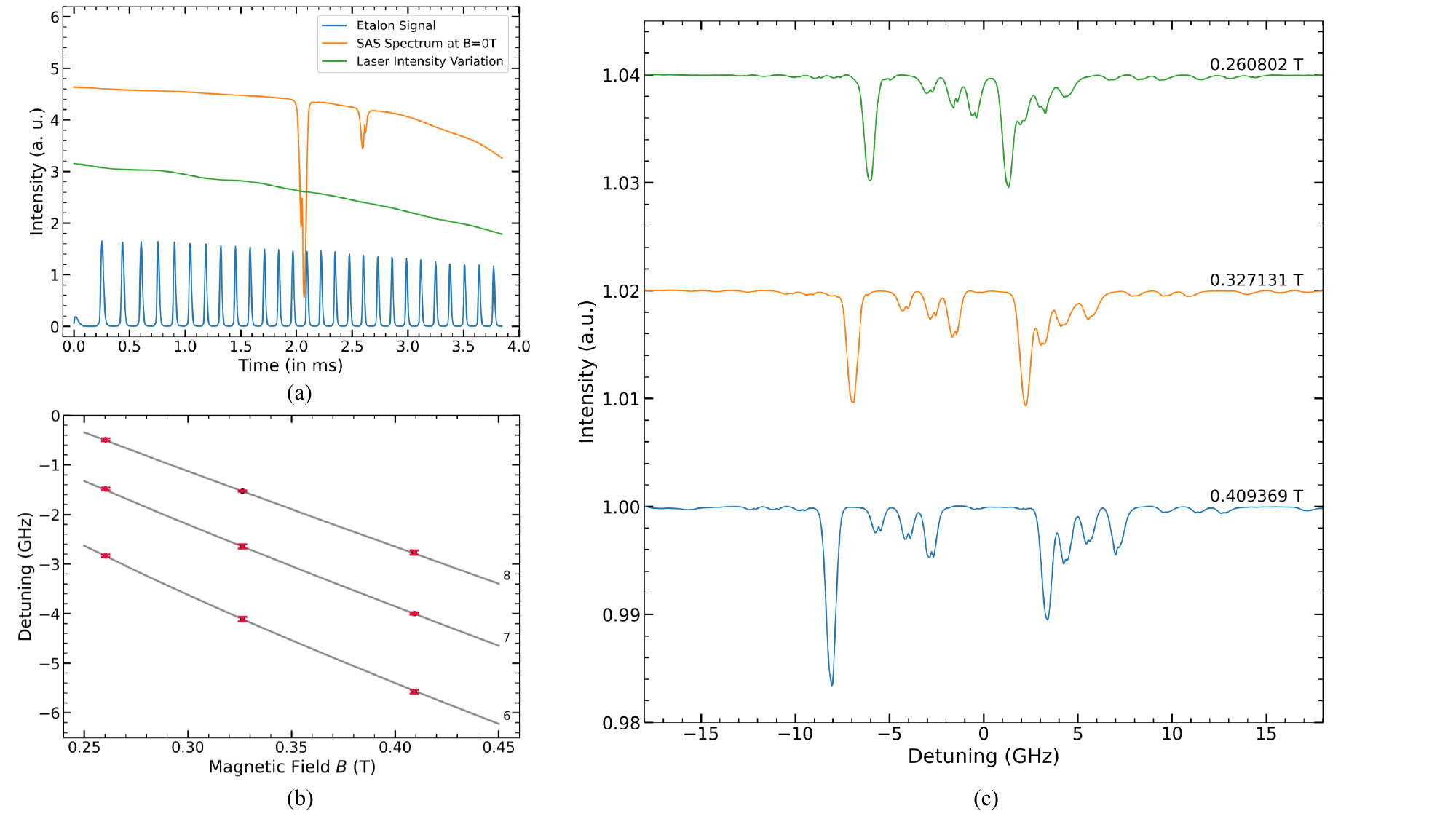}
\caption{\label{fig:MeasuredSpectra}(a) Frequency calibration and reference signals. The blue trace shows the Fabry–Perot transmission peaks (detected by PD3). The orange trace is the saturated absorption spectrum (SAS) of the $^{87}\text{Rb}$ $D_2$ line at zero field (PD2). The green trace (PD4) monitors the change in laser intensity during the ramping of the laser. (b) Calculated $\sigma-$ transition detunings are shown as a function of the applied magnetic field $B$ (gray solid lines). Red circles indicate experimentally extracted resonance positions at $B = 0.260802$ T, $0.327131$ T, and $0.409369$ T, with vertical error bars representing the combined $1\sigma$ uncertainty from frequency calibration and peak fitting. Transitions 6, 7, and 8 belong to the $\ket{m_J=-1/2}\rightarrow\ket{m_J'=-3/2}$ manifold with $m_I = -1/2, +1/2,$ and $+3/2$, respectively. (c) Experimental spectra obtained for the $^{87}\text{Rb}$ $D_2$ transition at magnetic fields of $0.260802\,\text{T}$, $0.327131\,\text{T}$, and $0.409369\,\text{T}$ (from top to bottom).}
\end{figure*}

The light source is a Photodigm Distributed Bragg Reflector (DBR) Laser, tuned to the $\mathrm{D}_2$ transition line of ${}^{87}\text{Rb}$ ($\lambda \approx 780.241 \text{ nm}$). An Optical Isolator (OI) prevents back-reflections from reaching the laser cavity. The output beam is initially split into two paths. A half-wave plate and a Polarizing Beam Splitter (PBS) are used to direct the light into the reference arm and the main experiment arm. The light source provides a mode-hop-free spectral scan range of approximately $35 \text{ GHz}$. This wide range is crucial, as it enables the simultaneous capture of multiple widely separated atomic transitions characteristic of the Hyperfine Paschen-Back regime.

The reference arm serves two essential functions for frequency calibration. First, a portion of the reference beam is directed through a Thorlabs Fabry-Perot Interferometer to produce frequency markers for coarse frequency axis calibration. Second, another portion of the reference beam is directed into a conventional reference SAS setup containing a separate ${}^{87}\text{Rb}$ cell at zero field. This zero-field spectrum is used for calibrating the zero detuning position of the high-field SAS spectrum.

The main experiment arm measures the ${}^{87}\text{Rb}$ spectrum under a strong magnetic field. The sensor head consists of isotopically pure $^{87}\mathrm{Rb}$ contained within a $12 \text{ mm} \times 12 \text{ mm} \times 12 \text{ mm}$ vapor cell from Precision Glassblowing Inc. An external longitudinal magnetic field ranging from $0.2 \text{ T}$ to $~0.4 \text{ T}$ is generated using two permanent Neodymium (N42) ring magnets. Each magnet has an inner diameter of $0.25 \text{ inches}$, outer diameter of $2 \text{ inches}$, and a thickness of $1 \text{ inch}$, and they are configured in a quasi-Helmholtz arrangement. The basic magnetic housing design was adapted from the magnetic field configuration scheme described in Ref.~\cite{reed2018low}. This specific configuration was engineered to withstand the substantial attractive forces between the two permanent magnets as well as achieve a uniform field distribution across the vapor cell. The vapor cell is in a Faraday geometry, where the direction of light propagation is parallel to the applied magnetic field. In this configuration, the linearly polarized probe decomposes into $\sigma^+$ and $\sigma^-$ components, driving electric-dipole transitions with $\Delta m_J = \pm 1$ and $\Delta m_I=0$ in the uncoupled $|m_I, m_J\rangle$ basis. To stabilize the atomic vapor density and enhance the overall signal-to-noise ratio ($\text{SNR}$), the cell is actively heated and maintained at a temperature of $\sim 37^\circ \text{C}$.

\subsection{Spectral Measurements}
The sensor response was characterized at three magnetic field strengths (0.2--0.4~T) by mechanically adjusting the permanent magnet separation. For each configuration, the raw photodetector signal is converted to a calibrated spectrum using the reference optical assembly illustrated in Fig.~\ref{fig:MeasuredSpectra}(a).
% Accurate field estimation requires precise knowledge of the frequency axis. The calibration of the Data Acquisition (DAQ) system's time axis to the frequency detuning scale is established using the reference optical assembly illustrated in Fig. \ref{fig:MeasuredSpectra} (a). 
The etalon serves as a precise frequency ruler, and the reference SAS peak at $0\,T$ provides the absolute frequency reference. To construct the frequency axis, the etalon signal from PD3 undergoes preprocessing with a Savitzky-Golay filter, which preserves the etalon peaks while eliminating high-frequency noise. Subsequently, a peak detection algorithm from SciPy is employed to identify the coarse peaks. To surpass the resolution limit imposed by the Data Acquisition (DAQ) system sampling rate, the precise center of each etalon peak was determined via a sub-pixel parabolic fit to the local maxima. This procedure yields a set of time indices corresponding to frequency markers separated by $1.5\,\text{GHz}$. The full continuous frequency axis, $\nu(t)$, was generated by linearly interpolating between the identified peaks, thereby correcting for any non-linearities in the laser-scanning mechanism. Finally, the absolute detuning $\Delta=0$ tied to the unperturbed $^{87}\text{Rb}$ $5S_{1/2} \rightarrow 5P_{3/2}$ transition energy, determined from the reference SAS at 0 T (Fig. \ref{fig:MeasuredSpectra} (a)). 

The calibrated spectra at fields of 0.260802~T, 0.327131~T, and 0.409369~T are shown in Fig.~\ref{fig:MeasuredSpectra}(b). With increasing magnetic field, the $\sigma^+$ and $\sigma^-$ manifolds diverge, giving a Doppler-free spectral fingerprint. Each manifold contains eight resolved sub-Doppler peaks corresponding to the eight ground-state sublevels ($m_I$, $m_J=\pm 1/2$) coupled by $\sigma^\pm$ transitions to the allowed excited states. The monotonic dependence of the transition frequencies on the applied field, combined with the sub-Doppler resolution, gives the basis for the magnetic field estimation described in the following subsection.

\subsection{\label{ssec:2C}Magnetic Field Estimation and Sensitivity}
Quantitative estimation of the magnetic field is performed via a physics-constrained spectral optimization  \cite{scotto2016rubidium}. 
Saturated-absorption peaks $\boldsymbol{\nu}_{\text{exp}}^{(k)}$ were extracted by locally fitting each resolved SAS feature with an inverted Gaussian background plus a skewed Lorentzian line shape. The Lorentzian center was used as the transition frequency, with the $1\sigma$ uncertainty determined from the fit covariance matrix.
% Saturated absorption peaks $\boldsymbol{\nu}_{\text{exp}}^{(k)}$ are extracted by fitting Gaussian profiles to each resolved SAS feature in the experimental spectrum.
These positions are compared against the theoretical transition frequencies, $\nu_{theo}(B)$, derived from the diagonalization of the Hamiltonian $H$ described in eq. \ref{eq:Hamiltonian}. The estimated field is the value of $B$ that minimizes the least-squares residual ($L(B)$):
\begin{equation}
    L(B) = \sum_{k} \left( \nu_{exp}^{(k)} - \nu_{theo}^{(k)}(B) \right)^2
\end{equation}
A bounded minimization routine returns the optimum $B$ for each measured spectrum. 

Measurement uncertainty is propagated through this inversion by Monte Carlo (MC). We model each experimental peak position $\nu_{exp}^{(k)}$ as an independent random variable following a Gaussian distribution with standard deviation $\sigma_{k}$. The total uncertainty for each peak is calculated by combining the statistical fitting error ($\sigma_{fit}^{(k)}$) and the global frequency axis calibration uncertainty ($\sigma_{calib}$) in quadrature:
\begin{equation} 
    \sigma_{k} = \sqrt{(\sigma_{fit}^{(k)})^2 + (\sigma_{calib})^2} 
\end{equation}

We generate $N_{MC}$ synthetic datasets by perturbing the measured peak positions according to:
\begin{equation}
\nu_{\mathrm{exp,MC}}^{(k,j)} = \nu_{\mathrm{exp}}^{(k)} + \delta_{k,j}, \qquad \quad \delta_{k,j} \sim \mathcal{N}(0, \sigma_k)
\end{equation}
For each MC trial $j$, the optimization routine minimizes $L(B)$ to yield a trial field estimate $B_j$. The final reported magnetic field uncertainty is defined as the standard deviation of the resulting distribution of estimated fields $\{B_j\}_{j=1}^{N_{MC}}$. 

Using this technique, the magnetic field was estimated for the three experimental configurations shown in Fig.~\ref{fig:MeasuredSpectra}. The analysis yielded field values of $0.409369 \pm 0.000561\,\text{T}$, $0.327131 \pm 0.000668\,\text{T}$, and $0.260802 \pm 0.000746\,\text{T}$. 
% The consistent per-shot precision across the dynamic range indicates that the field-retrieval procedure is well-conditioned throughout this range.

The sensitivity is computed from the single-shot field uncertainty and the spectrum acquisition time. Each spectrum is acquired by averaging 8 consecutive laser scans of $7.69$ ms each, for a total integration time ($T_{int}$) of 61.5 ms per field estimate. Combined with the single-shot precision of approximately $\sigma_B\approx$ $561\, \mu$T, this gives a noise-equivalent magnetic field $\sigma_B\sqrt{T_{int}}\approx$  $139\,\mu\text{T}/\sqrt{\text{Hz}}$ at $\sim 0.4 \, \text{T}$. This sensitivity is currently limited by the uncertainty in the frequency axis calibration, the finite sampling resolution of the DAQ system, and the relatively low laser scan rate.

% \begin{figure*}[t]
% \centering
% % --- First row (single figure) ---
% \subfigure[]{%
%     \includegraphics[width=0.8\textwidth]{img/Optical_Setup.pdf}
%     \label{fig:2a}}

% \vspace{1em} % vertical space between rows

% % --- Second row (two figures) ---
% \subfigure[]{%
%     \includegraphics[width=0.38\textwidth]{fig_1.eps}
%     \label{fig:2b}}
% \hfill
% \subfigure[]{%
%     \includegraphics[width=0.38\textwidth]{fig_1.eps}
%     \label{fig:2c}}

% \caption{(a) Caption for subfigure 1. 
% (b) Caption for subfigure 2. 
% (c) Caption for subfigure 3.}
% \label{fig:wide}
% \end{figure*}

\section{\label{sec:level3}Theoretical Modeling}

The field estimation procedure described in Sec.~\ref{sec:level2} relies solely on the measured peak positions. However, the experimental spectra contain additional information in their line strengths, lineshapes, and relative amplitudes, all of which depend on laser power, temperature, polarization, and the degree of state mixing in the Paschen-Back regime. A model that captures these features is valuable both for interpreting the complex spectral structure and for generating synthetic training data toward future autonomous field estimation via machine learning. Therefore, we develop a multilevel optical Bloch equation (OBE) model for $^{87}$Rb saturated absorption spectroscopy in the uncoupled $\ket{m_I, m_J}$ basis, and validate it against the experimental spectra presented above.

\subsection{Hyperfine Paschen-Back Hamiltonian and Basis Selection}
Our modeling spans the transition from the Zeeman regime to the hyperfine Paschen-Back regime (HPB). For $^{87}$Rb in intermediate-to-high fields, the Zeeman interaction exceeds the hyperfine coupling, so the total angular momentum $F$ is no longer conserved, and the eigenstates must be described in an uncoupled basis ($\ket{m_I, m_J}$). We numerically diagonalize the full Hamiltonian ($H$):
\begin{equation}
    \label{eq:Hamiltonian}
    H = H_0 + H_{hfs} + H_B
\end{equation}
where $H_0$ is the unperturbed atomic Hamiltonian, $H_{hfs}$ is the hyperfine interaction term, and $H_B$ is the Zeeman interaction term describing the coupling to the external magnetic field $B$. The numerical solution yields the precise eigenvalues of the states and the eigenstates as the linear combination of uncoupled basis states $\ket{m_I, m_J}$.

% \begin{figure*}[t]
%     \label{fig:intro}
%     \centering
%     % --- First row (two figures) ---
%     \subfigure[]{%
%         \includegraphics[width=0.47\textwidth, trim={0 -2.75cm 0 0}]{img/FlowChart.png}
%         \label{fig:ModelingFlowChart}}
%     \hfill
%     \subfigure[]{%
%         \includegraphics[width=0.51\textwidth]{img/SimulatedResult_combined.png}
%         \label{fig:SimulatedSpectra}}
%     % \vspace{1em}
%     % \centering
%     % % --- First row (single figure) ---
%     % \subfigure[]{%
%     %     \includegraphics[width=\textwidth]{img/SimulatedResult_0p4092T.png}
%     %     \label{fig:SimulatedSAS_0p4092T}}
    
%     % \vspace{1em} % vertical space between rows
%     \caption{(a) Computational workflow for modeling high-field SAS using an optical Bloch equation (OBE) framework. (b) Simulated spectra obtained for the $^{87}\text{Rb}$ $D_2$ transition at magnetic fields of $0.2602\,\text{T}$, $0.3263\,\text{T}$, and $0.4092\,\text{T}$ using the Optical Bloch Equation (OBE) model at the same field strengths. All traces are vertically offset for clarity.}
% \end{figure*}

\subsection{Saturated Absorption Spectroscopy Modeling}
We model the high-field SAS by extending the density-matrix framework of Maguire et al. \cite{Maguire_Bijnen_Mese_Scholten_2006}, originally developed for $^{85}$Rb in a zero field in the $\ket{F, m_F}$ basis. The high-field regime requires a structural reformulation rather than a parameter change. In the hyperfine Paschen-Back regime, $F$ is no longer a good quantum number, so the natural basis for both the eigenstates and dipole matrix elements is the uncoupled $|m_I$, $m_J\rangle$ basis; transition frequencies $\omega_{\beta\alpha}$ are computed from the field-dependent eigenstates by diagonalizing Eq.~\ref{eq:Hamiltonian} and the angular coupling coefficients $C_{\beta\alpha}$ are evaluated via the transformation of the bare-basis coupling matrix into the eigenstate basis (see Appendix~\ref{app:coupling}).
We retain the master-equation structure of Maguire et al. \cite{Maguire_Bijnen_Mese_Scholten_2006} but solve it on the field-dependent eigenstate manifold of the $D_2$ transition ($5^2S_{1/2} \rightarrow 5^2P_{3/2}$), which involves a total of $N=24$ magnetic sublevels. The indices $\alpha = \{1, 2, ... ,8\}$ label the ground-state sublevels of the $5^2S_{1/2}$ manifold, while $\beta = \{9, 10, ..., 24\}$ label the excited-state sublevels of the $5^2P_{3/2}$ manifold. This model incorporates the magnetic field-dependent atomic Hamiltonian, polarization-selective transitions, and Doppler averaging based on a Maxwell-Boltzmann velocity distribution. 

The density-matrix evolution in the rotating frame can be written as (with derivation provided in Appendix~A):
\begin{equation}\label{eq:obe_ground}
    \begin{aligned}
        \dot{\rho}_{\alpha\alpha} \!=\!2\cos{(kz)}\!\sum_{\beta=9}^{24} \Omega_{\beta\alpha} \mathrm{Im}(\tilde{\rho}_{\beta\alpha})\!+\!\frac{1}{\tau}\!\sum_{\beta=9}^{24} (C_{\beta\alpha})^{2}\!\rho_{\beta\beta}
    \end{aligned}
\end{equation}
\begin{equation}\label{eq:obe_excited}
    \begin{aligned}
        \dot{\rho}_{\beta\beta}= -2\cos(kz)\sum_{\alpha=1}^{8}\Omega_{\beta\alpha}\mathrm{Im}\big(\tilde{\rho}_{\beta\alpha}\big)-\frac{1}{\tau}\rho_{\beta\beta}
    \end{aligned}
\end{equation}
\begin{equation}\label{eq:obe_coherence}
    \begin{aligned}
        \dot{\tilde{\rho}}_{\beta\alpha}= i\Omega_{\beta\alpha}\cos{\!(kz)}(\rho_{\beta\beta}-\rho_{\alpha\alpha})-\!i(\omega_{\beta\alpha}-\!\omega)\tilde{\rho}_{\beta\alpha}-\\\frac{1}{2\tau}\tilde{\rho}_{\beta\alpha} \qquad (\text{for }\alpha \neq \beta)
    \end{aligned}
\end{equation}
Here, $\rho_{\alpha\alpha}$ and $\rho_{\beta\beta}$ represent the populations of the ground and excited states, respectively, while $\tilde{\rho}_{\beta\alpha}$ denotes the slowly varying optical coherence term. The quantity $\omega_{\beta\alpha}$ is the magnetic-field-dependent atomic transition frequency, $\omega$ is the laser angular frequency, $k$ is the optical wave vector, and $\tau$ is the excited-state lifetime.
The $C_{\beta\alpha}$ represents angular coupling coefficients for the transition from $\beta$ to $\alpha$ and the rabi frequency ($\Omega_{\beta\alpha}$) is defined as \cite{Maguire_Bijnen_Mese_Scholten_2006}:
\begin{equation} \label{eq:rabifreq}
    \Omega_{\beta\alpha} = C_{\beta\alpha}\Bigg(\frac{3\pi \epsilon_0}{k^3\tau \hbar}\Bigg)^{\frac{1}{2}}E
\end{equation}
where $E$ is the electric field amplitude of the traveling wave, $\epsilon_0$ is vacuum permittivity and $\hbar$ is reduced Planck constant.

\begin{figure}[t]
    \includegraphics[width=0.47\textwidth, trim={0 0 0 0}]{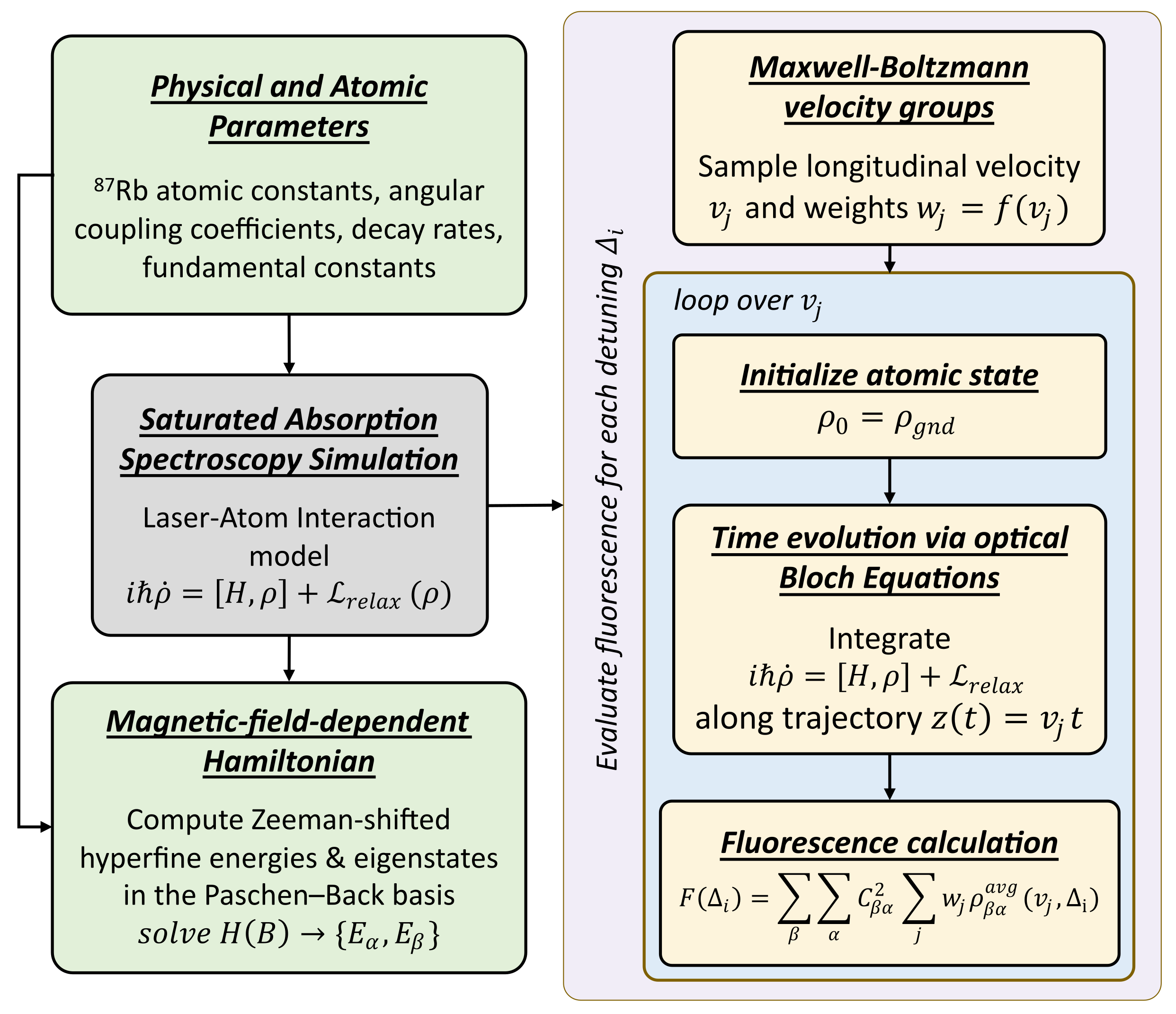}
    \caption{\label{fig:ModelingFlowChart} Computational workflow for modeling high-field SAS using an optical Bloch equation (OBE) framework.}
\end{figure}

The full description of the atomic system involves an $N \times N$ matrix of coupled equations (Eqs. (\ref{eq:obe_ground}) - (\ref{eq:obe_coherence})), where $N=24$. 
% is the total number of magnetic sublevels involved in $D_2$ transition $(5^2S_{1/2} \rightarrow 5^2P_{3/2})$. 
This results in a total of 576 equations, which is computationally demanding to solve directly. However, the system can be significantly simplified by considering selection rules and inherent symmetries of the problem. First, the hermiticity of the density matrix, $\rho_{\beta\alpha} = \rho_{\alpha\beta}^*$, reduces the problem to $N(N+1)/2$. Second, the model is significantly simplified by the fact that the optical field, $H^I_{\beta\alpha}$, does not drive coherences between levels within the same manifold. Consequently, all ground–ground and excited–excited coherences vanish, meaning only the diagonal population elements within these manifolds need to be retained. Furthermore, we restrict the light–atom interaction term $H^I_{\beta\alpha}$ to the transitions permitted by the dipole selection rules ($\Delta m_J = \pm 1$) determined by the laser polarization. While state mixing in the high-field and intermediate regimes can give rise to ``magnetically induced" (MI) transitions that are strictly forbidden at zero field \cite{tonoyan2018circular, sargsyan2021circular, zentile2014hyperfine}, these transitions remain secondary to the principal dipole transitions in the field range under investigation. The inclusion of MI transitions would enhance the model’s theoretical completeness, but their omission facilitates a substantial reduction in computational complexity. Implementing these physical and mathematical constraints reduces the original 576 equations to a manageable set of 40 coupled differential equations.

\subsection{Numerical Solution and Calculation of SAS Signal}

We applied the Runge-Kutta method to solve the system of coupled differential equations. The density matrix $\rho(v, \Delta, t)$ gives the number of atoms in each of their internal states at time $t$ for a certain velocity $v$ and laser detuning $\Delta$. Fig. \ref{fig:ModelingFlowChart} illustrates the computational workflow used to obtain the numerical solution.

\begin{figure*}[t]
\centering
% {left bottom right top}
\includegraphics[width=\textwidth, trim={0.55cm 1.45cm 0.5cm 0.5cm}, clip]{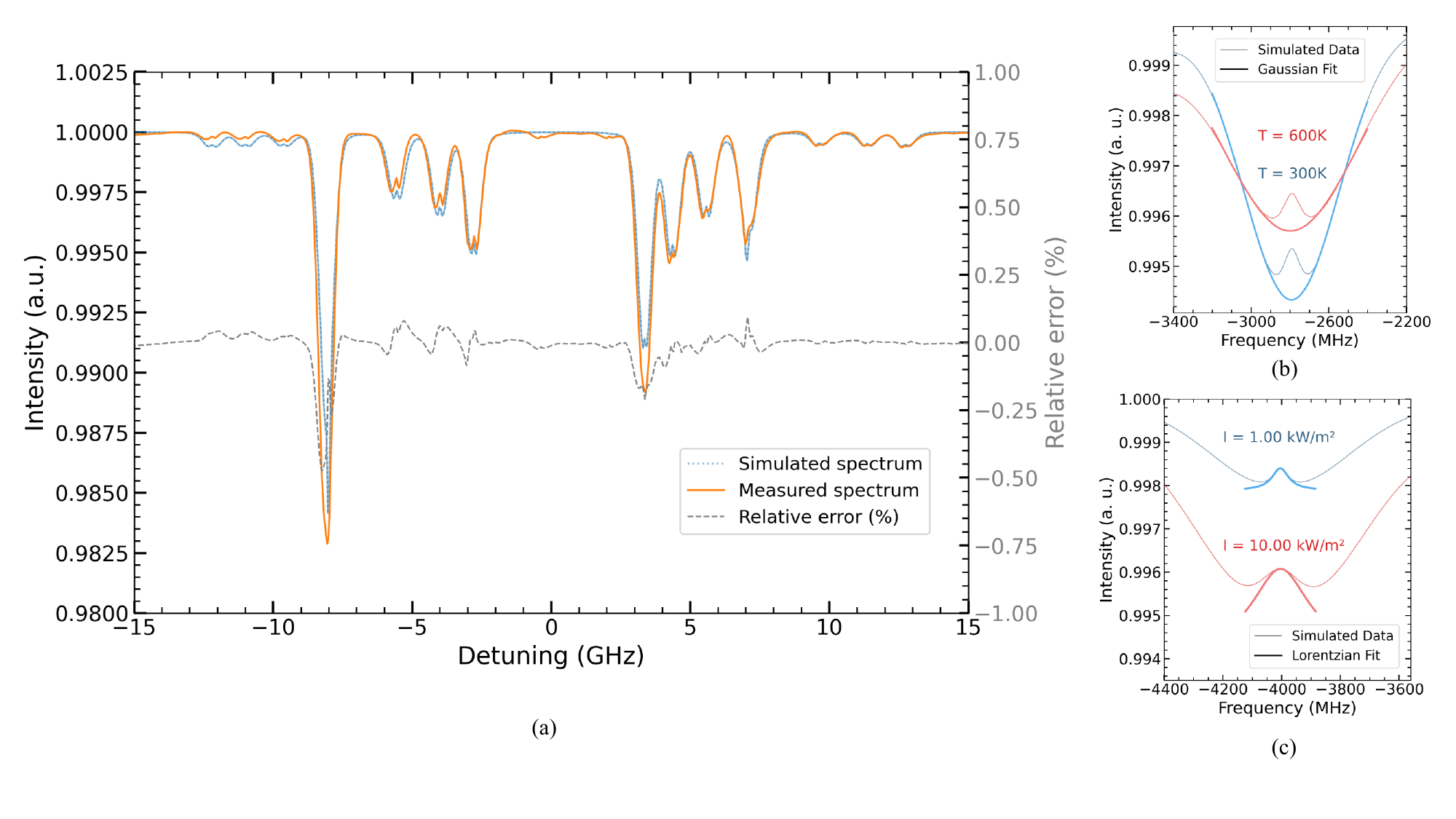}

\caption{\label{fig:ModelingValidation}(a) Simulated spectra overlayed with experimental spectra obtained for the $^{87}\text{Rb}$ $D_2$ transition at magnetic field of $0.409369\,\text{T}$. (b) Doppler-broadened profiles at $T = 300\,\text{K}$ (blue) and $600\,\text{K}$ (red). (c) Simulated spectra showing power broadening at $I = 1.0\,\text{kW/m}^2$ and $10.0\,\text{kW/m}^2$ at $T = 310\,\text{K}$.}
\end{figure*}

The observed SAS spectrum represents an effective absorption signal that is derived from the total fluorescence emitted by an ensemble of atoms. This fluorescence is averaged over time and the Maxwellian velocity distribution. The fluorescence resulting from the decay of a single excited state, denoted as $\beta$, is expressed as \cite{Maguire_Bijnen_Mese_Scholten_2006}:

\begin{equation}
F_{\beta} = \sum_{\alpha} C^2_{\beta\alpha}\rho_{\beta\beta}
\end{equation}

The total fluorescence signal $F(\Delta)$, which represents the ensemble-averaged absorption measured at a given laser detuning $\Delta$, is calculated by summing the contributions from all excited states $\beta$ ($\sum_{\beta} F_{\beta}$) after performing the velocity-averaging step:

\begin{equation}
    F(\Delta) = \sum_\beta\sum_{\alpha} C^2_{\beta\alpha} \sum_if(v_i)\rho^{\text{avg}}_{\beta\beta}(v_i, \Delta)
\end{equation}

Here, $f(v_i)$ is the Maxwellian probability for velocity group $v_i$, and $\rho^{\text{avg}}_{\beta\beta}(v_i, \Delta)$ is the time-averaged excited-state population calculated from the OBE for that velocity group. The detuning $\Delta$ is defined as the frequency difference between the laser frequency $\omega_L$ and the centered frequency $\omega_0$ of the $5^2S_{1/2} \to 5^2P_{3/2}$ fine structure transition, $\Delta = \omega_L - \omega_0$.

\subsection{Model Validation}
The OBE model was validated against the experimental spectra using simulation parameters constrained to match the experimental conditions: a probe beam waist ($w_0$) of $0.84\,\text{mm}$, an incident pump power of $5.5\,\text{mW}$, and a temperature of $310$K ($37^o$C). The atomic properties, including transition strengths, natural lifetimes, and hyperfine splitting constants, were obtained from the standard reference for $^{87}\text{Rb}$ $D$ line data \cite{steck_rubidium_2025}.

Figure~\ref{fig:ModelingValidation}(a) shows the simulated spectrum overlaid with the experimental spectrum at B$= 0.409369 \ \text{T}$, along with the relative error between the two. The model reproduces the spectral fingerprint with accurate relative positions and absorption depth across both manifolds. The relative error between the simulated and experimental plot is between $\pm0.25\%$ for most of the spectral range. The larger deviation across the strong absorption feature could be due to the experimental noise, baseline drift, or other experimental conditions such as variation of the temperature of the vapor cell and laser intensity noise. The simulated spectra at all measured field strengths show similar agreement. % ($0.2602 \ \text{T}$, $0.3263 \ \text{T}$, and $0.4092\ \text{T}$) are shown in Fig.~\ref{fig:SimulatedSpectra}, validating that the model successfully predicts the resonance positions and the relative absorption depths.
Beyond peak positions, the OBE model captures the fundamental broadening mechanisms inherent to the system:
\begin{itemize}
    \item \textbf{Doppler Broadening:} The model takes into account of longitudinal Maxwell-Boltzmann velocity distribution $f(v, T)$. As illustrated in Fig. \ref{fig:ModelingFlowChart}, each velocity class is treated independently: for every velocity group, the optical Bloch equations are solved with the corresponding Doppler-shifted detuning, and the resulting responses are thermally averaged over the distribution. To validate this thermal averaging, we simulated the spectral envelope at $T = 300\,\text{K}$ and $T = 600\,\text{K}$ of the 
    $\ket{\nicefrac{3}{2}, \nicefrac{-1}{2}} \rightarrow \ket{\nicefrac{3}{2}, \nicefrac{-3}{2}}$
    % $\ket{3/2, -1/2} \rightarrow \ket{3/2, -3/2}$ 
    transition (see Fig. \ref{fig:ModelingValidation}(b).  The extracted FWHM values were found to be $578.31 \pm 3.31\,\text{MHz}$ and $812.59 \pm 24.21\,\text{MHz}$, respectively. The ratio of these widths is $0.712$, which stands in excellent agreement with the theoretical scaling ratio of $\sqrt{\nicefrac{300}{600}} \approx 0.707$ expected for a Doppler-dominated profile. This confirms that the simulation accurately captures the characteristic $\sqrt{T}$ dependence of the inhomogeneous broadening.
    \item \textbf{Power Broadening:} The atom-light interaction strength is governed by the Rabi frequency $\Omega$ defined in Eq. \ref{eq:rabifreq}, which scales with the dipole coupling $\Omega \propto \mathbf{d} \cdot \mathbf{E}$, leading to intensity-dependent spectral broadening. To characterize this effect and validate the model's saturation parameters, we analyzed the lineshape of the $\ket{\nicefrac{1}{2}, \nicefrac{-1}{2}} \rightarrow \ket{\nicefrac{1}{2}, \nicefrac{-3}{2}}$ transition at two distinct intensities: $I_{1} = 1.0\,\text{kW/m}^2$ and $I_{2} = 10.0\,\text{kW/m}^2$ (see Fig. \ref{fig:ModelingValidation}(c)). Lorentzian fits were applied to the simulated data profiles to extract the Full Width at Half Maximum (FWHM). The ratio of the measured intensity broadening, $\Delta\omega_{FWHM}(I_{1})/\Delta\omega_{FWHM}(I_{2})$, was determined to be 0.302. This value agrees extremely well with the theoretical ratio of 0.318 obtained from the standard power-broadening expression derived for a driven two-level system, $\Delta\omega_{FWHM} = \Gamma \sqrt{1 + \nicefrac{I}{I_{sat}}}$ \cite{foot2005atomic}, confirming that the OBE model accurately captures the non-linear saturation dynamics of the system.
\end{itemize}

The simulation effectively predicts the shape of the spectrum, as well as the effects of Doppler and power broadening. However, the model does not consider collisional broadening, since its contribution is small compared to Doppler and power broadening at the temperature maintained in the vapor cell \cite{weller2011absolute}. The model considers changes in the velocity distribution with temperature; however, it does not account for the increase in atomic number density that is governed by the vapor pressure of rubidium. As a result, absolute signal amplitudes are treated as normalized parameters.

\section{\label{sec:level4}Discussion and Conclusion}
In this paper, we demonstrate SASHMAG, an all-optical magnetometer that resolves the sub-Doppler Zeeman transitions of ${}^{87}Rb$ in the hyperfine Paschen-Back regime and extracts the magnetic field via a physics-constrained optimization.
This multi-transition approach provides redundancy against spectral noise and does not require prior knowledge of the magnetic field. Our approach yielded consistently precise field extraction across the dynamic range from $0.2$ to $0.4~$T (limited by our DC magnets) with an uncertainty of $\pm 0.14\%$. 
SASHMAG instead resolves the full Zeeman-split $\mathrm{D_2}$ manifold in a single laser sweep, trading peak sensitivity for absolute field retrieval without prior knowledge of the field.
% The cesium EXAAQ magnetometer of St${\ae}$rkind et al. \cite{staerkind2023precision, staerkind2024high} achieves high sensitivity at $7~T$ by tracking a single extreme-state transition, but requires the probe laser to be pre-positioned near the expected field-dependent transition using a sideband modulation chain locked to a zero-field reference cell. Adapting this to variable-field applications such as fusion would require repositioning the probe laser near the expected transition and reconfiguring the modulation chain for each new field. 
Recent high-field Cs magnetometer work \cite{staerkind2023precision, staerkind2024high, guo2026tesla} achieves higher resolution than SASHMAG, but does so via frequency-modulation techniques at a fixed operating point. Extending to a continuous field range would require retuning the laser and modulation chain to track the shifting transitions. SASHMAG instead resolves the full Zeeman-split $D_2$ manifold in a single laser sweep, trading peak resolution for absolute field retrieval across a continuous range without prior knowledge of the field.

% The cesium EXAAQ magnetometer of St${\ae}$rkind et al. \cite{staerkind2023precision, staerkind2024high} demonstrates that alkali-vapor magnetometry can achieve good sensitivity at multi-tesla fields by continuously tracking a single extreme-state transition using sideband spectroscopy with the 5th harmonic of a $\sim 19.5\,\text{GHz}$ modulator to bridge the large frequency offset between the zero-field reference and the field-dependent transition. The two approaches address different measurement scenarios. EXAAQ architecture is optimized for high sensitivity at a known operating point, but extension to variable-field applications such as fusion would require re-positioning the probe laser near the expected transition and reconfiguring the modulation chain for each new field. 

We also validated a multilevel optical Bloch equation (OBE) model in the hyperfine Paschen-Back regime, capturing the saturation effects and spectral structure that arise in this high-power, high-field environment. We confirmed the model's accuracy by comparing it against fundamental non-linear physics: the observed power-broadening ratio and the Doppler-broadening thermal scaling matched our analytical predictions closely.
The current sensitivity of $139\,\mu\text{T}/\sqrt{\text{Hz}}$ at $\sim 0.4 \, \text{T}$ is primarily limited by the laser scan rate and frequency calibration uncertainty. To assess the achievable sensitivity floor of this approach, we applied the same optimization procedure to peak positions extracted from noise-free simulated spectra. Assuming a laser scan rate of $250\,\text{Hz}$ with four averaging cycles, this yields a projected sensitivity of $8\,\mu\text{T}/\sqrt{\text{Hz}}$, indicating that over an order of magnitude improvement is accessible through faster scanning and improved frequency calibration alone, without changes to the measurement principle.
Despite this, the all-optical architecture, involving no electronic components at the sensor head, makes SASHMAG a promising candidate for deployment in harsh environments where conventional sensors face fundamental limitations.

\section{\label{sec:level6}Outlook}
Future development of SASHMAG will focus on enhancing the sensor's performance and achieving full system autonomy. To increase both measurement sensitivity and bandwidth, optical interrogation schemes involving electro-optic modulation can be used \cite{staerkind2024high}, achieving high bandwidth to capture transient magnetic field fluctuations relevant to MHD detection in the plasma environments. Since the OBE model validated in this work accurately captures field-dependent SAS lineshapes, it can be used to generate synthetic training data for machine-learning approaches to peak identification and field interference. This would enable real-time field estimation without human-supervised peak fitting, a natural direction for autonomous deployment \cite{li2020intelligent}.

\begin{acknowledgments}
This work was supported by the US Department of Energy (DE-SC0024471). This work utilized computational resources provided by the Center for High Throughput Computing (CHTC) at the University of Wisconsin--Madison. The authors acknowledge the use of CHTC services supported by the University of Wisconsin--Madison\cite{chtcuwmadison}. We also acknowledge the University of Wisconsin--Madison Design and Innovation Lab for assistance with the fabrication of experimental components.\\

% \textbf{Author contributions:}
% M.D.
\end{acknowledgments}

\appendix
\section{Derivation of Optical Bloch Equations}
The time evolution of the density matrix elements, derived from the Liouville-von Neumann equation, is given by:
\begin{equation}\label{eq:rho}
    \dot{\rho_{\beta\alpha}} = -i \omega_{\beta\alpha}\rho_{\beta\alpha} - \frac{i}{\hbar}\sum_\gamma\big(H^{I}_{\beta\gamma}\rho_{\gamma\alpha} - \rho_{\beta\gamma}H^{I}_{\gamma\alpha}\big)
\end{equation}
Phenomenological decay terms are added to account for spontaneous emission and decoherence:
\begin{align}
    -\frac{1}{2\tau}\,\rho_{\beta\alpha}
    &\qquad \text{for }\dot{\rho}_{\beta\alpha}\; (\beta\neq\alpha) \\[6pt]
    -\frac{1}{\tau}\,\rho_{\beta\beta}
    &\qquad \text{for }\dot{\rho}_{\beta\beta} \\[6pt]
    +\frac{1}{\tau}\sum_{\beta} (C_{\beta\alpha})^{2}\,\rho_{\beta\beta}
    &\qquad \text{for }\dot{\rho}_{\alpha\alpha}
    \end{align}
Here, the ground states are defined by $\alpha \in \{1, \ldots, 8\}$ and the excited states by $\beta \in \{9, \ldots, 24\}$.

Following the transformation of the off-diagonal elements to a rotating frame defined by the atom-laser coupling frequency $\omega$, we define the tilde matrix elements:
\begin{equation}\label{eq:rho_tilde_RFTransf}
    \tilde{\rho}_{\beta\alpha} =
    \begin{cases}
    \rho_{\beta\alpha}\, e^{i\omega t} & (\beta \ne \alpha),\\
    \rho_{\beta\alpha} & (\beta = \alpha).
    \end{cases}
\end{equation}

\subsection{Ground-State Diagonals ($\dot{\rho}_{\alpha\alpha}$)}
For the time evolution of the ground-state diagonals, $\dot{\rho}_{\alpha\alpha}$, using equation \eqref{eq:rho} and including the phenomenological decay and repopulation terms:
\begin{equation}
    \dot{\rho_{\alpha\alpha}} = - \frac{i}{\hbar}\sum_\gamma\big(H^{I}_{\alpha\gamma}\rho_{\gamma\alpha} - \rho_{\alpha\gamma}H^{I}_{\gamma\alpha}\big) +\frac{1}{\tau}\sum_{\beta} (C_{\beta\alpha})^{2}\,\rho_{\beta\beta}
\end{equation}
Since the interaction Hamiltonian $H^{I}$ couples only between ground states ($\alpha$) and excited states ($\beta$), the summation over $\gamma$ reduces to only the excited states $\beta$:
\begin{equation}\label{eq:rho_aa_full}
    \dot{\rho_{\alpha\alpha}} =  - \frac{i}{\hbar}\sum_\beta\big(H^{I}_{\alpha\beta}\rho_{\beta\alpha} - \rho_{\alpha\beta}H^{I}_{\beta\alpha}\big) +\frac{1}{\tau}\sum_{\beta} (C_{\beta\alpha})^{2}\,\rho_{\beta\beta}
\end{equation}
The interaction Hamiltonian matrix element is defined as:
\begin{equation}
    H^{I}_{\beta\alpha} = \hbar\Omega_{\beta\alpha}\hat{E}
\end{equation}
where $\hat{E} = \cos{(kz)}(e^{i\omega t}+e^{-i\omega t})$. 
% The term $H^{I}_{\alpha\beta}$ is the Hermitian conjugate of $H^{I}_{\beta\alpha}$, such that $H^{I}_{\alpha\beta} = (H^{I}_{\beta\alpha})^* = \hbar\Omega_{\alpha\beta}^*\hat{E}$. For real Rabi frequencies $\Omega_{\beta\alpha} = \Omega_{\alpha\beta}$, and 
Since $\rho_{\alpha\beta} = \rho_{\beta\alpha}^*$, the first term in Equation \eqref{eq:rho_aa_full} simplifies as follows:
\begin{equation}
    \begin{aligned}
    - \frac{i}{\hbar} \sum_\beta\big(H^{I}_{\alpha\beta} & \rho_{\beta\alpha}
   - \rho_{\alpha\beta}  H^{I}_{\beta\alpha}\big)\\
    &= 
    - \frac{i}{\hbar}\sum_\beta\big(H^{I}_{\alpha\beta}\rho_{\beta\alpha} - \rho_{\alpha\beta}H^{I}_{\beta\alpha}\big)\\
    &= -\frac{i}{\hbar}\sum_\beta\big(\hbar\Omega_{\alpha\beta}\hat{E}\rho_{\beta\alpha} - \rho_{\alpha\beta}\hbar\Omega_{\beta\alpha}\hat{E}\big)\\
    &=-i \sum_\beta \Omega_{\beta\alpha} \big(\hat{E}\rho_{\beta\alpha} - \rho_{\alpha\beta}\hat{E}\big)
    \end{aligned}
\end{equation}
\begin{equation}
    \begin{aligned}
    - \frac{i}{\hbar} \sum_\beta\big(&H^{I}_{\alpha\beta} \rho_{\beta\alpha}
   - \rho_{\alpha\beta}  H^{I}_{\beta\alpha}\big)\\
    &= -i \sum_\beta \Omega_{\beta\alpha} \big(\hat{E}\rho_{\beta\alpha} - \rho_{\alpha\beta}\hat{E}\big)\\
    &= -i \cos{(kz)}\sum_{\beta}\Omega_{\beta\alpha}(e^{i\omega t}+e^{-i\omega t})(\rho_{\beta\alpha}-\rho^*_{\beta\alpha})
    \end{aligned}
\end{equation}
Next, we substitute $\rho_{\beta\alpha} = \tilde{\rho}_{\beta\alpha}e^{-i\omega t}$ and apply the Rotating Wave Approximation (RWA), which drops terms evolving at $e^{\pm 2i\omega t}$:
\begin{equation}
    \begin{aligned}
        - \frac{i}{\hbar} & \sum_\beta\big(H^{I}_{\alpha\beta} \rho_{\beta\alpha}
   - \rho_{\alpha\beta}  H^{I}_{\beta\alpha}\big)\\
    &= -i \cos{(kz)}\sum_{\beta}\Omega_{\beta\alpha}(e^{i\omega t}+e^{-i\omega t})(\rho_{\beta\alpha}-\rho^*_{\beta\alpha})\\
    &= -i \cos{(kz)}\sum_{\beta}\Omega_{\beta\alpha}(e^{i\omega t}+e^{-i\omega t})(\tilde{\rho}_{\beta\alpha}e^{-i\omega t}-\tilde{\rho}^*_{\beta\alpha}e^{i\omega t})\\
    &=2\cos{(kz)}\sum_\beta \Omega_{\beta\alpha} \mathrm{Im}(\tilde{\rho_{\beta\alpha}})
    \end{aligned}
\end{equation}
Substituting this simplified term back into Equation \eqref{eq:rho_aa_full} yields the final equation for the ground-state diagonals:
\begin{equation}
    \dot{\rho_{\alpha\alpha}} =2\cos{(kz)}\sum_\beta \Omega_{\beta\alpha} \mathrm{Im}(\tilde{\rho_{\beta\alpha}}) +\frac{1}{\tau}\sum_{\beta} (C_{\beta\alpha})^{2}\,\rho_{\beta\beta}
\end{equation}

% Excited State Diagonals
\subsection{Excited-State Diagonals ($\dot{\rho}_{\beta\beta}$)}

For the time evolution of the excited-state diagonals, we start from the master
equation,
\begin{equation}
    \dot{\rho}_{\beta\beta}
    = -\frac{i}{\hbar}\sum_\gamma\!\left(
        H^{I}_{\beta\gamma}\rho_{\gamma\beta}
        - \rho_{\beta\gamma}H^{I}_{\gamma\beta}
    \right)
    -\frac{1}{\tau}\rho_{\beta\beta},
\end{equation}
and note that the interaction Hamiltonian couples only excited states
$\beta$ to ground states $\alpha$. Thus, the sum reduces to
\begin{equation}\label{eq:rho_bb_reduced}
    \dot{\rho}_{\beta\beta} =
    -\frac{i}{\hbar}\sum_{\alpha}
        \left(
            H^{I}_{\beta\alpha}\rho_{\alpha\beta}
            -\rho_{\beta\alpha}H^{I}_{\alpha\beta}
        \right)
    -\frac{1}{\tau}\rho_{\beta\beta}.
\end{equation}

Using the definition of the interaction Hamiltonian,
$H^{I}_{\beta\alpha} = \hbar\Omega_{\beta\alpha}\,\hat{E}$ with
$\hat{E}(z,t)=\cos(kz)\big(e^{i\omega t}+e^{-i\omega t}\big)$, the first term becomes
\begin{equation}
    \begin{aligned}
        -\frac{i}{\hbar}\!\sum_\alpha\!\big(&
        H^{I}_{\beta\alpha} \rho_{\alpha\beta}
        -\rho_{\beta\alpha}H^{I}_{\alpha\beta}
    \big)\\
    &=
    -i\cos(kz)\sum_\alpha\Omega_{\beta\alpha}
        (e^{i\omega t}+e^{-i\omega t})
        (\rho_{\alpha\beta}-\rho_{\beta\alpha}). 
    \end{aligned}
\end{equation}

We now transform to the rotating frame using the definition of eq. \ref{eq:rho_tilde_RFTransf}
\begin{equation*}
    \rho_{\beta\alpha} = \tilde{\rho}_{\beta\alpha}e^{-i\omega t},
\qquad
    \rho_{\alpha\beta} = \tilde{\rho}_{\beta\alpha}^* e^{i\omega t}.
\end{equation*}
Substituting these and applying the RWA gives,
\begin{equation*}
    (e^{i\omega t}+e^{-i\omega t})
    (\rho_{\alpha\beta}-\rho_{\beta\alpha})
    = \tilde{\rho}_{\beta\alpha}^* - \tilde{\rho}_{\beta\alpha}
    = -2i\,\mathrm{Im}(\tilde{\rho}_{\beta\alpha})
\end{equation*}

Thus, the coherent contribution becomes
\begin{equation}
    \begin{aligned}
        -i\cos(kz)\sum_\alpha\Omega_{\beta\alpha} &
    \big(\tilde{\rho}_{\beta\alpha}^* 
    -\tilde{\rho}_{\beta\alpha}\big)\\
    &=
    -2\cos(kz)\sum_\alpha \Omega_{\beta\alpha}\,
    \mathrm{Im}\!\left(\tilde{\rho}_{\beta\alpha}\right).
    \end{aligned}
\end{equation}

Substituting this into Eq.~\eqref{eq:rho_bb_reduced}, we obtain the final form
for the excited-state diagonal elements:
\begin{equation}
    \dot{\rho}_{\beta\beta}
    = -\,2\cos(kz)
        \sum_{\alpha}\Omega_{\beta\alpha}\,
        \mathrm{Im}\!\big(\tilde{\rho}_{\beta\alpha}\big)
    -\frac{1}{\tau}\rho_{\beta\beta}
\end{equation}

\subsection{Off-Diagonal Elements ($\dot{\rho}_{\beta\alpha}$, $\beta\neq\alpha$)}

Equation \ref{eq:rho} gives us the time evolution of the off-diagonal density-matrix elements, which includes terms for phenomenological coherence decay.

For general indices, if $\beta$ is not equal to $\alpha$,
\begin{equation}\label{eq:rho_offdiag}
    \dot{\rho}_{\beta\alpha}
    = -i\omega_{\beta\alpha}\rho_{\beta\alpha}
    -\frac{i}{\hbar}\sum_{\gamma}\!\big(H^{I}_{\beta\gamma}\rho_{\gamma\alpha}
    - \rho_{\beta\gamma}H^{I}_{\gamma\alpha}\big)
    -\frac{1}{2\tau}\rho_{\beta\alpha}
\end{equation}
where $\omega_{\beta\alpha}=\omega_\beta-\omega_\alpha$.
Considering the rotating frame transformation as described in Eq. \ref{eq:rho_tilde_RFTransf} we have:
\begin{equation}
    \tilde{\rho}_{\beta\alpha} = \rho_{\beta\alpha}\, e^{i\omega t}
\end{equation}
differentiating both sides w.r.t time
\begin{equation}
    \dot{\tilde{\rho}}_{\beta\alpha} = \dot{\rho}_{\beta\alpha}\, e^{i\omega t} 
    + i\omega\rho_{\beta\alpha}\, e^{i\omega t}
\end{equation}
Rearranging the terms will give us,
\begin{equation}\label{eq:rho_tilde_rho_dot_tilde}
    \dot{\rho}_{\beta\alpha} = (\dot{\tilde{\rho}}_{\beta\alpha}-i\omega \tilde{\rho}_{\beta\alpha})e^{-i\omega t}
\end{equation}
Substituting eq. \ref{eq:rho_tilde_rho_dot_tilde} into eq. \ref{eq:rho_offdiag} 

\begin{equation}
    \begin{aligned}
       (\dot{\tilde{\rho}}_{\beta\alpha}-i\omega &\tilde{\rho}_{\beta\alpha})e^{-i\omega t}
    = -i\omega_{\beta\alpha}\rho_{\beta\alpha}\\&
    -\frac{i}{\hbar}\sum_{\gamma}\!\big(H^{I}_{\beta\gamma}\rho_{\gamma\alpha}
    - \rho_{\beta\gamma}H^{I}_{\gamma\alpha}\big)
    -\frac{1}{2\tau}\rho_{\beta\alpha} 
    \end{aligned}
\end{equation}
\begin{equation}
    \begin{aligned}
       \dot{\tilde{\rho}}_{\beta\alpha}-i\omega \tilde{\rho}_{\beta\alpha}
    =& -i\omega_{\beta\alpha}\displaystyle{\underbrace{\rho_{\beta\alpha}e^{i\omega t}}_{{\tilde{\rho}}_{\beta\alpha}}} -\frac{1}{2\tau}\displaystyle{\underbrace{\rho_{\beta\alpha}e^{i\omega t}}_{{\tilde{\rho}}_{\beta\alpha}}} \\&
    -\frac{i}{\hbar}\sum_{\gamma}\!\big(H^{I}_{\beta\gamma}\displaystyle{\underbrace{\rho_{\gamma\alpha}e^{i\omega t}}_{{\tilde{\rho}}_{\gamma\alpha}}}
    - \displaystyle{\underbrace{\rho_{\beta\gamma}e^{i\omega t}}_{{\tilde{\rho}}_{\beta\gamma}}}H^{I}_{\gamma\alpha}\big)    
    \end{aligned}
\end{equation}
\begin{equation}
    \begin{aligned}
       \dot{\tilde{\rho}}_{\beta\alpha}
    = -i(\omega_{\beta\alpha}-&\omega)\tilde{\rho}_{\beta\gamma}-\frac{1}{2\tau}\tilde{\rho}_{\beta\gamma}\\&
    -\frac{i}{\hbar}\sum_{\gamma}\!\big(H^{I}_{\beta\gamma}\tilde{\rho}_{\gamma\alpha}
    - \tilde{\rho}_{\beta\gamma}H^{I}_{\gamma\alpha}\big)
    \end{aligned}
\end{equation}

Given that the laser is in resonance with the $\alpha \leftrightarrow \beta$ transition. In the initial term of the summation, the dominant contribution occurs when $\gamma = \alpha$, yielding: $\sum_{\gamma}H^{I}_{\beta\gamma}\tilde{\rho}_{\beta\gamma} \approx H^{I}_{\beta\alpha}\tilde{\rho}_{\alpha\alpha}$. Similarly, for the subsequent term: The dominant contribution occurs when $\gamma = \beta$, giving $\sum_{\gamma}H^{I}_{\beta\gamma}\tilde{\rho}_{\beta\gamma} \approx H^{I}_{\beta\alpha}\tilde{\rho}_{\beta\beta}$.
Thus, 
\begin{equation}
    \dot{\tilde{\rho}}_{\beta\alpha}= i\Omega_{\beta\alpha}\cos{\!(kz)}(\rho_{\beta\beta}-\rho_{\alpha\alpha})-\!i(\omega_{\beta\alpha}-\!\omega)\tilde{\rho}_{\beta\alpha}-\\\frac{1}{2\tau}\tilde{\rho}_{\beta\alpha}
\end{equation}

\subsection{\label{app:coupling}Angular Coupling Coefficients}

The angular dependence of the electric-dipole interaction is obtained using
the Wigner--Eckart theorem. The Clebsch--Gordan coefficient can be written in
terms of Wigner's 3$j$ symbol {\cite{farrell1995rabi}} as

\begin{equation}
\label{eq:Cq}    
    \begin{aligned}
        \langle J_1J_2m_1m_2|J_1J_2Jm\rangle = 
(-1)^{-J_1+J_2-m}& \\
\sqrt{2J+1}
\begin{pmatrix}
J_1 & J_2 & J \\
m_1 & m_2 & -m
\end{pmatrix}
    \end{aligned}
\end{equation}

For a transition driven by a photon with polarization $q=0,\pm1$, the corresponding angular coupling coefficient is

\begin{equation}
C^{(q)}_{m_J',m_J}
=
(-1)^{J'-1+m_J}
\sqrt{2J+1}
\begin{pmatrix}
J' & 1 & J \\
m_J' & q & -m_J
\end{pmatrix},
\end{equation}

which is nonzero only when $m_J'=m_J+q$ and $m_{I}^{'}=m_{I}$.

The coupling matrix $C^{(q)}$ is first constructed in the
$|m_J,m_I\rangle$ basis using the Wigner 3-$j$ symbol (Eq.~\ref{eq:Cq}). However, the magnetic-field eigenstates are superpositions of these basis states, so the coupling matrix in the eigenstate basis is obtained through the transformation
\begin{equation}
C_{\mathrm{eig}}^{(q)} = U_g^\dagger C^{(q)} U_e
\end{equation}
where $U_g$ and $U_e$ are the eigenvector matrices of the ground-
and excited-state Hamiltonians, respectively.

% whether or not they are actually referenced in the text. This is an appropriate
% \nocite{*}

\bibliography{apssamp}% Produces the bibliography via BibTeX.

@PREAMBLE{
 "\providecommand{\noopsort}[1]{}" 
 # "\providecommand{\singleletter}[1]{#1}%" 
}

@article{staerkind2023precision,
  title={Precision measurement of the excited state land{\'e} g-factor and diamagnetic shift of the cesium d 2 line},
  author={St{\ae}rkind, Hans and Jensen, Kasper and M{\"u}ller, J{\"o}rg H and Boer, Vincent O and Petersen, Esben T and Polzik, Eugene S},
  journal={Physical Review X},
  volume={13},
  number={2},
  pages={021036},
  year={2023},
  publisher={APS}
}

@article{sargsyan2015study,
  title={Study of the Rb D 2-line splitting in a strong transverse magnetic field with Doppler-free spectroscopy in a nanocell},
  author={Sargsyan, A and Hakhumyan, G and Tonoyan, A and Petrov, PA and Vartanyan, TA},
  journal={Optics and Spectroscopy},
  volume={119},
  number={2},
  pages={202--207},
  year={2015},
  publisher={Springer}
}

@article{reed2018low,
  title={Low-drift Zeeman shifted atomic frequency reference},
  author={Reed, DJ and {\v{S}}ibali{\'c}, N and Whiting, DJ and Kondo, JM and Adams, CS and Weatherill, KJ},
  journal={OSA Continuum},
  volume={1},
  number={1},
  pages={4--12},
  year={2018},
  publisher={Optical Society of America}
}

@article{Maguire_Bijnen_Mese_Scholten_2006, title={Theoretical calculation of saturated absorption spectra for multi-level atoms}, volume={39}, DOI={10.1088/0953-4075/39/12/007}, number={12}, journal={Journal of Physics B: Atomic, Molecular and Optical Physics}, author={Maguire, L P and Bijnen, R M and Mese, E and Scholten, R E}, year={2006}, month={May}, pages={2709–2720}}

@article{staerkind2024high,
  title={High-field optical cesium magnetometer for magnetic resonance imaging},
  author={St{\ae}rkind, Hans and Jensen, Kasper and M{\"u}ller, J{\"o}rg H and Boer, Vincent O and Polzik, Eugene S and Petersen, Esben T},
  journal={PRX Quantum},
  volume={5},
  number={2},
  pages={020320},
  year={2024},
  publisher={APS}
}

@article{zapf2014bose,
  title={Bose-Einstein condensation in quantum magnets},
  author={Zapf, Vivien and Jaime, Marcelo and Batista, CD},
  journal={Reviews of Modern Physics},
  volume={86},
  number={2},
  pages={563--614},
  year={2014},
  publisher={APS}
}

@article{lewin2023review,
  title={A review of UTe2 at high magnetic fields},
  author={Lewin, Sylvia K and Frank, Corey E and Ran, Sheng and Paglione, Johnpierre and Butch, Nicholas P},
  journal={Reports on Progress in Physics},
  volume={86},
  number={11},
  pages={114501},
  year={2023},
  publisher={IOP Publishing}
}

@article{kartsovnik2004high,
  title={High magnetic fields: a tool for studying electronic properties of layered organic metals},
  author={Kartsovnik, Mark V},
  journal={Chemical reviews},
  volume={104},
  number={11},
  pages={5737--5782},
  year={2004},
  publisher={ACS Publications}
}

@misc{ong_quantum_2021,
	title = {Quantum matter in ultrahigh magnetic fields},
	url = {http://arxiv.org/abs/2103.09155},
	doi = {10.48550/arXiv.2103.09155},
	urldate = {2025-12-18},
	publisher = {arXiv},
	author = {Ong, N. P. and Li, Lu},
	month = mar,
	year = {2021},
	note = {arXiv:2103.09155 [cond-mat]},
}

@article{bottura2022superconducting,
  title={Superconducting magnets and technologies for future colliders},
  author={Bottura, Luca and Prestemon, Soren and Rossi, Lucio and Zlobin, Alexander V},
  journal={Frontiers in Physics},
  volume={10},
  pages={935196},
  year={2022},
  publisher={Frontiers Media SA}
}

@article{shen2022design,
  title={Design, fabrication, and characterization of a high-field high-temperature superconducting Bi-2212 accelerator dipole magnet},
  author={Shen, Tengming and Garcia Fajardo, Laura and Myers, Cory and Hafalia Jr, Aurelio and Rudeiros Fern{\'a}ndez, Jose Luis and Arbelaez, Diego and Brouwer, Lucas and Caspi, Shlomo and Ferracin, Paolo and Gourlay, Stephen and others},
  journal={Physical Review Accelerators and Beams},
  volume={25},
  number={12},
  pages={122401},
  year={2022},
  publisher={APS}
}

@article{muon2021magnetic,
  title={Magnetic-field measurement and analysis for the Muon g-2 Experiment at Fermilab},
  author={Muon g-2 Collaboration and others},
  journal={Physical Review A},
  volume={103},
  number={4},
  pages={042208},
  year={2021}
}

@article{ma_design_2016,
	title = {Design and development of {ITER} high-frequency magnetic sensor},
	volume = {112},
	issn = {0920-3796},
	url = {https://www.sciencedirect.com/science/article/pii/S0920379616303374},
	doi = {10.1016/j.fusengdes.2016.05.002},
	urldate = {2025-12-18},
	journal = {Fusion Engineering and Design},
	author = {Ma, Y. and Vayakis, G. and Begrambekov, L. B. and Cooper, J. -J. and Duran, I. and Hirsch, M. and Laqua, H. P. and Moreau, Ph. and Oosterbeek, J. W. and Spuig, P. and Stange, T. and Walsh, M.},
	month = nov,
	year = {2016},
	pages = {594--612},
}

@article{moreau_new_2018,
	title = {The new magnetic diagnostics in the {WEST} tokamak},
	volume = {89},
	issn = {0034-6748},
	url = {https://doi.org/10.1063/1.5036537},
	doi = {10.1063/1.5036537},
	number = {10},
	urldate = {2025-12-18},
	journal = {Review of Scientific Instruments},
	author = {Moreau, P. and Le-Luyer, A. and Spuig, P. and Malard, P. and Saint-Laurent, F. and Artaud, J. F. and Morales, J. and Faugeras, B. and Heumann, H. and Cantone, B. and Moreau, M. and Brun, C. and Nouailletas, R. and Nardon, E. and Santraine, B. and Berne, A. and Kumari, P. and Belsare, S. and {WEST Team}},
	month = jul,
	year = {2018},
	pages = {10J109},
}

@article{vayakis_development_2012,
	title = {Development of the {ITER} magnetic diagnostic set and specificationa)},
	volume = {83},
	issn = {0034-6748},
	url = {https://doi.org/10.1063/1.4732077},
	doi = {10.1063/1.4732077},
	number = {10},
	urldate = {2025-12-18},
	journal = {Review of Scientific Instruments},
	author = {Vayakis, G. and Arshad, S. and Delhom, D. and Encheva, A. and Giacomin, T. and Jones, L. and Patel, K. M. and Pérez-Lasala, M. and Portales, M. and Prieto, D. and Sartori, F. and Simrock, S. and Snipes, J. A. and Udintsev, V. S. and Watts, C. and Winter, A. and Zabeo, L.},
	month = jul,
	year = {2012},
	pages = {10D712},
}

@article{kraff2015mri,
  title={MRI at 7 Tesla and above: demonstrated and potential capabilities},
  author={Kraff, Oliver and Fischer, Anja and Nagel, Armin M and M{\"o}nninghoff, Christoph and Ladd, Mark E},
  journal={Journal of Magnetic Resonance Imaging},
  volume={41},
  number={1},
  pages={13--33},
  year={2015},
  publisher={Wiley Online Library}
}

@article{garn1966technique,
  title={Technique for measuring megagauss magnetic fields using Zeeman effect},
  author={Garn, WB and Caird, RS and Thomson, DB and Fowler, CM},
  journal={Review of Scientific Instruments},
  volume={37},
  number={6},
  pages={762--767},
  year={1966}
}

@article{gomez2014magnetic,
  title={Magnetic field measurements via visible spectroscopy on the Z machine},
  author={Gomez, MR and Hansen, SB and Peterson, KJ and Bliss, DE and Carlson, AL and Lamppa, DC and Schroen, DG and Rochau, GA},
  journal={Review of Scientific Instruments},
  volume={85},
  number={11},
  pages={11E609},
  year={2014}
}

@article{banasek2016measuring1,
  title={Measuring 10--20 T magnetic fields in single wire explosions using Zeeman splitting},
  author={Banasek, JT and Engelbrecht, JT and Pikuz, SA and Shelkovenko, TA and Hammer, DA},
  journal={Review of Scientific Instruments},
  volume={87},
  number={10},
  pages={103506},
  year={2016}
}

@article{banasek2016measuring2,
  title={Measuring 20--100 T B-fields using Zeeman splitting of sodium emission lines on a 500 kA pulsed power machine},
  author={Banasek, JT and Engelbrecht, JT and Pikuz, SA and Shelkovenko, TA and Hammer, DA},
  journal={Review of Scientific Instruments},
  volume={87},
  number={11},
  pages={11D407},
  year={2016}
}

@article{george2017pulsed,
  title={Pulsed high magnetic field measurement with a rubidium vapor sensor},
  author={George, S and Bruyant, N and B{\'e}ard, J and Scotto, S and Arimondo, E and Battesti, R and Ciampini, D and Rizzo, C},
  journal={Review of Scientific Instruments},
  volume={88},
  number={7},
  pages={073102},
  year={2017}
}

@article{ciampini2017optical,
  title={Optical spectroscopy of a microsized Rb vapor sample in magnetic fields up to 58 T},
  author={Ciampini, D and Battesti, R and Rizzo, C and Arimondo, E},
  journal={Physical Review A},
  volume={96},
  number={5},
  pages={052504},
  year={2017}
}

@article{klinger2020proof,
  title={Proof of the feasibility of a nanocell-based wide-range optical magnetometer},
  author={Klinger, E and Azizbekyan, H and Sargsyan, A and Leroy, C and Sarkisyan, D and Papoyan, A},
  journal={Applied Optics},
  volume={59},
  number={7},
  pages={2231--2237},
  year={2020}
}

@article{keaveney2018elecsus,
  title={ElecSus: Extension to arbitrary geometry magneto-optics},
  author={Keaveney, James and Adams, Charles S and Hughes, Ifan G},
  journal={Computer physics communications},
  volume={224},
  pages={311--324},
  year={2018},
  publisher={Elsevier}
}

@article{zentile2015elecsus,
  title={ElecSus: A program to calculate the electric susceptibility of an atomic ensemble},
  author={Zentile, Mark A and Keaveney, James and Weller, Lee and Whiting, Daniel J and Adams, Charles S and Hughes, Ifan G},
  journal={Computer Physics Communications},
  volume={189},
  pages={162--174},
  year={2015},
  publisher={Elsevier}
}

@article{trattnig2018key,
  title={Key clinical benefits of neuroimaging at 7 T},
  author={Trattnig, Siegfried and Springer, Elisabeth and Bogner, Wolfgang and Hangel, Gilbert and Strasser, Bernhard and Dymerska, Barbara and Cardoso, Pedro Lima and Robinson, Simon Daniel},
  journal={Neuroimage},
  volume={168},
  pages={477--489},
  year={2018},
  publisher={Elsevier}
}

@article{okada2022neuroimaging,
  title={Neuroimaging at 7 Tesla: a pictorial narrative review},
  author={Okada, Tomohisa and Fujimoto, Koji and Fushimi, Yasutaka and Akasaka, Thai and Thuy, Dinh HD and Shima, Atsushi and Sawamoto, Nobukatsu and Oishi, Naoya and Zhang, Zhilin and Funaki, Takeshi and others},
  journal={Quantitative imaging in medicine and surgery},
  volume={12},
  number={6},
  pages={3406},
  year={2022}
}

@phdthesis{scotto2016rubidium,
  title={Rubidium vapors in high magnetic fields},
  author={Scotto, Stefano},
  year={2016},
  school={Universit{\'e} Paul Sabatier-Toulouse III}
}

@misc{chtcuwmadison,
  doi = {10.21231/GNT1-HW21},
  url = {https://chtc.cs.wisc.edu/},
  author = {{Center for High Throughput Computing}},
  title = {Center for High Throughput Computing},
  publisher = {Center for High Throughput Computing},
  year = {2006}
}

@misc{steck_rubidium_2025,
  author = {Daniel A. Steck},
  title = {Rubidium 87 D Line Data},
  howpublished = {available online at \url{http://steck.us/alkalidata}},
  year = {2025},
  note = {Revision 2.3.4}
}

@book{foot2005atomic,
  title={Atomic physics},
  author={Foot, Christopher J},
  volume={7},
  year={2005},
  publisher={Oxford university press}
}

@article{buzio2011fabrication,
  title={Fabrication and calibration of search coils},
  author={Buzio, Marco},
  journal={arXiv preprint arXiv:1104.0803},
  year={2011}
}

@article{graham2026high,
  title={High Field Diamond Magnetometry Towards Tokamak Diagnostics},
  author={Graham, SM and Stephen, CJ and Newman, AJ and Edmonds, AM and Markham, ML and Morley, GW},
  journal={arXiv preprint arXiv:2601.13413},
  year={2026},
  url = "https://arxiv.org/pdf/2601.13413"
}

@article{zentile2014hyperfine,
  title={The hyperfine Paschen--Back Faraday effect},
  author={Zentile, Mark A and Andrews, Rebecca and Weller, Lee and Knappe, Svenja and Adams, Charles S and Hughes, Ifan G},
  journal={Journal of Physics B: Atomic, Molecular and Optical Physics},
  volume={47},
  number={7},
  pages={075005},
  year={2014},
  publisher={IOP Publishing}
}

@article{tonoyan2018circular,
  title={Circular dichroism of magnetically induced transitions for D2 lines of alkali atoms},
  author={Tonoyan, A and Sargsyan, A and Klinger, E and Hakhumyan, G and Leroy, C and Auzinsh, M and Papoyan, A and Sarkisyan, D},
  journal={Europhysics Letters},
  volume={121},
  number={5},
  pages={53001},
  year={2018},
  publisher={EDP Sciences, IOP Publishing and Societ{\`a} Italiana di Fisica}
}

@article{sargsyan2021circular,
  title={Circular dichroism in atomic vapors: magnetically induced transitions responsible for two distinct behaviors},
  author={Sargsyan, Armen and Amiryan, Arevik and Tonoyan, Ara and Klinger, Emmanuel and Sarkisyan, David},
  journal={Physics Letters A},
  volume={390},
  pages={127114},
  year={2021},
  publisher={Elsevier}
}

@incollection{boivin2016diagnostics,
  title={Diagnostics for magnetic fusion power plants},
  author={Boivin, R},
  booktitle={Magnetic Fusion Energy},
  pages={549--575},
  year={2016},
  publisher={Elsevier}
}

@article{bolshakova2012hall,
  title     = {Magnetic measuring instrumentation with radiation-resistant {Hall} sensors for fusion reactors: Experience of testing at {JET}},
  author    = {Bolshakova, I. and Quercia, A. and Pironti, A. and Holyaka, R. and Duran, I. and Murari, A. and {JET Contributors}},
  journal   = {IEEE Transactions on Nuclear Science},
  volume    = {59},
  number    = {4},
  pages     = {1224--1231},
  year      = {2012},
  doi       = {10.1109/TNS.2012.2198671},
  publisher = {IEEE}
}

@article{quercia2022longterm,
  title     = {Long term operation of the radiation-hard {Hall} probes system and the path toward a high performance hybrid magnetic field sensor},
  author    = {Quercia, A. and Pironti, A. and Bolshakova, I. and Holyaka, R. and Duran, I. and Murari, A. and {JET Contributors}},
  journal   = {Nuclear Fusion},
  volume    = {62},
  number    = {10},
  pages     = {106032},
  year      = {2022},
  doi       = {10.1088/1741-4326/ac8a03},
  publisher = {IOP Publishing}
}

@misc{lakeshore_teslameter,
  title        = {{F71} and {F41} teslameter specifications},
  author       = {{Lake Shore Cryotronics, Inc.}},
  howpublished = {\url{https://www.lakeshore.com/products/categories/specification/magnetic-products/gaussmeters-teslameters/f71-and-f41-teslameters}},
  note         = {Accessed: 2025-12-26},
  year         = {2025}
}

@article{tsujilio2001fiberoptic,
  title     = {Fiberoptic heterodyne magnetic field sensor for long-pulsed fusion devices},
  author    = {Tsuji-Lio, S. and Akiyama, T. and Sato, E. and Nozawa, T. and Tsutsui, H. and Shimada, R. and Takahashi, M. and Terai, K.},
  journal   = {Review of Scientific Instruments},
  volume    = {72},
  number    = {1},
  pages     = {413--420},
  year      = {2001},
  doi       = {10.1063/1.1318248},
  publisher = {AIP Publishing}
}

@article{weller2011absolute,
  title={Absolute absorption on the rubidium D1 line including resonant dipole--dipole interactions},
  author={Weller, Lee and Bettles, Robert J and Siddons, Paul and Adams, Charles S and Hughes, Ifan G},
  journal={Journal of Physics B: Atomic, Molecular and Optical Physics},
  volume={44},
  number={19},
  pages={195006},
  year={2011}
}

@article{li2020intelligent,
  title={Intelligent and automatic laser frequency locking system using pattern recognition technology},
  author={Li, Qi-Xue and Zhang, Xu and Zhu, Ling-Xiao and Yan, Shu-Hua and Jia, Ai-Ai and Luo, Yu-Kun and Wang, Ya-Ning and Wei, Chun-Hua and Zhang, Huan-Kai and Lv, Meng-Jie and others},
  journal={Optics and Lasers in Engineering},
  volume={126},
  pages={105881},
  year={2020},
  publisher={Elsevier}
}

@article{farrell1995rabi,
  author  = {Farrell, P. M. and MacGillivray, W. R.},
  title   = {On the consistency of Rabi frequency calculations},
  journal = {Journal of Physics A: Mathematical and General},
  volume  = {28},
  pages   = {209--221},
  year    = {1995}
}

@article{guo2026tesla,
  title={Tesla-scale magnetic field measurement based on Sideband-overlap Zeeman spectroscopy using a functionalized MEMS vapor cell},
  author={Guo, Ju and Ma, Yintao and Lu, Dejiang and Yu, Mingzhi and Wang, Yanbin and Yang, Ping and Lin, Qijing and Zhao, Libo and Chen, Yao},
  journal={Microsystems \& Nanoengineering},
  volume={12},
  number={1},
  pages={219},
  year={2026},
  publisher={Nature Publishing Group}
}

@article{haupl2025modelling,
  title={Modelling spectra of hot alkali vapour in the saturation regime},
  author={H{\"a}upl, Daniel R and Higgins, Clare R and Pizzey, Danielle and Briscoe, Jack D and Wrathmall, Steven A and Hughes, Ifan G and L{\"o}w, Robert and Joly, Nicolas Y},
  journal={New Journal of Physics},
  volume={27},
  number={3},
  pages={033003},
  year={2025},
  publisher={IOP Publishing}
}

\end{document}